\documentclass[aps, twocolumn, nobibnotes, superscriptaddress, notitlepage, nobalancelastpage, noeprint, prb]{revtex4-2}
\usepackage[usenames,dvipsnames]{xcolor}
\usepackage[linktoc=page,colorlinks,urlcolor=RedViolet,citecolor=RedViolet,linkcolor=RedViolet]{hyperref}
\usepackage{float}
\usepackage{graphicx}
\usepackage{mathrsfs}
\usepackage{overpic}
\usepackage{wrapfig}
\usepackage{braket}
\usepackage{color}
\usepackage{verbatim}
\usepackage{amssymb,amsmath,mathtools}
\usepackage{upgreek}
\usepackage{bbold}
\usepackage{cancel}
\usepackage{textgreek}
\usepackage[normalem]{ulem}
\usepackage{mathtools}
\DeclarePairedDelimiter{\abs}{\lvert}{\rvert}
\definecolor{darkblue}{rgb}{0.0 0.0 0.78}
\definecolor{darkred}{rgb}{0.5 0.0 0.0}

\newcommand{\UMDphy}{Department of Physics, University of Maryland, College Park, Maryland 20742, USA}
\newcommand{\QTC}{Quantum Technology Center, University of Maryland, College Park, Maryland 20742, USA}
\newcommand{\UMDEECS}{Department of Electrical and Computer Engineering,
University of Maryland, College Park, Maryland 20742, USA}

\newcommand{\UMDBioE}{Fischell Department of Bioengineering, University of Maryland, College Park, Maryland 20742, USA}

\newcommand{\UMDBioChem}{Department of Chemistry and Biochemistry, University of Maryland, College Park, Maryland 20742, USA}

\begin{document}

\title{Eigenframe Synchronization in Disordered Driven Quantum Ensembles}
\date{\today}

\author{Saipriya Satyajit}
\affiliation{\UMDphy}
\affiliation{\QTC}
\author{Zechuan Yin}
\affiliation{\QTC}
\affiliation{\UMDEECS}
\author{Jner Tzern Oon}
\affiliation{\UMDphy}
\affiliation{\QTC}
\author{Katrijn Everaert}
\affiliation{\QTC}
\affiliation{\UMDEECS}
\author{Smriti Bhalerao}
\affiliation{\QTC}
\affiliation{\UMDBioE}
\author{John W. Blanchard}
\affiliation{\QTC}
\affiliation{\UMDEECS}
\author{Christopher Jarzynski}
\affiliation{\UMDphy}
\affiliation{\UMDBioChem}
\author{Ronald L. Walsworth}
\email{walsworth@umd.edu}
\affiliation{\UMDphy}
\affiliation{\QTC}
\affiliation{\UMDEECS}

\begin{abstract}
Periodic driving underlies many forms of quantum control, including for spin-based quantum sensing. However, in an
ensemble sensor, one global waveform must act on spins with different
detunings, drive amplitudes, hyperfine environments, and local fields.
Existing robust-control approaches to such disorder are usually framed in terms of
coherence, effective Hamiltonians, filter functions, or refocusing. Here
we identify a complementary geometric requirement for collective ensemble
Floquet control: disorder realizations must share a common Floquet
eigenframe. When this condition is met, initialization, protection, signal coupling, and readout are defined in one dressed basis, so the ensemble responds as a collective Floquet sensor rather than as an average over inequivalent driven members. We make this condition measurable with an eigenvector-based synchronization order parameter and a complementary fragmentation metric that quantify the alignment and spread of Floquet quantization axes across the ensemble. As one concrete realization of this framework, we use a continuous
counterdiabatic Floquet drive to derive synchronization criteria and
predict resonance-governed breakdown at the first two low-order
commensurabilities between the engineered Floquet gap and the drive
modulation, with detuning and amplitude disorder producing distinct
breakdown channels. Experiments on a nitrogen-vacancy (NV) ensemble in diamond verify the synchronized regime through long-lived collective oscillations, a two-dimensional disorder-robustness map, and breakdown resonances that shift with the modulation rate. Finally, we demonstrate a harmonic-free continuous-drive AC magnetometry protocol whose collective single-tone response is enabled by the synchronized Floquet eigenframe. These results establish Floquet eigenframe synchronization as a measurable condition for disorder-resilient collective control and quantum sensing.
\end{abstract}


\maketitle


\section{\label{sec:Intro}Introduction}


Periodic driving is a central tool for quantum control, enabling dynamical decoupling, dressed-state protection, Floquet engineering, and continuous quantum sensing~\cite{degen2017,goldman2014,cai2012}. However, in ensemble platforms for spin-based quantum sensing, global electromagnetic waveforms are typically used to initialize, protect, couple signals to, and read out many spins whose detunings, drive amplitudes, hyperfine environments, and local fields are not identical.
In such driven ensemble systems, robust-control strategies are commonly analyzed in terms of coherence preservation, average Hamiltonians~\cite{zhou2024prx}, filter functions~\cite{alvarez2011,bylander2011}, or refocusing of phase errors~\cite{souza2012}.
These perspectives are essential, but they do not by themselves answer a geometric question that becomes decisive in a driven ensemble: do all disorder realizations define the same Floquet basis?

The natural geometric objects in a periodically driven spin are the eigenvectors of the one-period propagator.
Their Bloch vectors define a Floquet eigenframe: the dressed quantization axes in which initialization, protection, signal coupling, and readout are interpreted.
Static disorder affects this Floquet structure in two distinct ways.
It can shift Floquet quasi-energies, producing phase dispersion that can sometimes be averaged or refocused~\cite{hahn1950,solomon1959,souza2012}.
It can also rotate the Floquet eigenvectors themselves, causing different ensemble members to occupy different dressed bases.
This second effect is a form of eigenframe fragmentation: even if each spin remains coherent in its own Floquet frame, the ensemble no longer behaves as one collective driven system.
Eigenvector rotation admits no analogous global inversion---the term that would need reversing points in a different direction for each member of the ensemble---and must instead be prevented by restoring a common effective frame over the relevant disorder range, the strategy implicit in composite pulses~\cite{levitt1986,cummins2003} and ensemble-compensating controls~\cite{likhaneja2006}.
Indeed, that the effective-field direction varies across an inhomogeneously broadened ensemble has been recognized since the earliest spin-locking experiments~\cite{redfield1955,hartmann1962} and is routinely managed spin by spin~\cite{garwood2001,laucht2017,seedhouse2021}; yet this axis dispersion has not, to our knowledge, been isolated as an ensemble-level object: no order parameter quantifies the alignment of Floquet quantization axes across a disordered ensemble, and no measured map delineates where a single global drive establishes---or fails to establish---a common Floquet eigenframe.

Here we identify the complementary geometric condition that determines whether a driven disordered ensemble responds collectively: the disorder realizations must share a common Floquet eigenframe.
We call this condition Floquet eigenframe synchronization.
The term synchronization is used here to mean alignment of Floquet quantization axes across an inhomogeneous ensemble, rather than dissipative phase locking~\cite{lee2013,walter2014,bai2024rydberg,schmolke2026}.
When synchronization holds, the ensemble has a common dressed basis and the collective response survives ensemble averaging.
When it fails, initialization, protection, signal coupling, and readout are defined in inequivalent local frames, and coherent single-spin dynamics are washed out by geometric inhomogeneity.
Synchronization is therefore what distinguishes an ensemble that behaves as a single collective Floquet object from one that behaves as an average over inequivalent driven spins.

We make this idea quantitative by introducing eigenvector-based metrics for a periodically driven ensemble.
For the adiabatic-following realization studied here, which we refer to as adiabatic frame synchronization (AFS), the synchronization order parameter $S_{\rm AFS}$ measures the collective alignment of Floquet quantization axes, while the complementary fragmentation metric $\sigma_{\rm AFS}$ measures their spread.
We show that synchronization produces three linked operational signatures: a common Floquet quantization axis, a common ensemble filter function, and a factorized collective population response.

We next consider implementation of AFS with a modulated drive that enforces adiabatic following at finite modulation rate and produces an engineered Floquet quasi-energy gap $\Omega_0$.
In the synchronized frame, amplitude and detuning disorder separate into diagonal quasi-energy shifts and off-diagonal eigenvector-mixing channels.
Synchronization is preserved when those mixing channels remain off-resonant from the engineered gap.

We find that the synchronized regime breaks down at discrete commensurability conditions where disorder-induced eigenvector-mixing terms become resonant with the engineered gap.
Floquet perturbation theory identifies the responsible disorder channels and predicts broad synchronized windows between these resonances, with a unified interpretation as avoided crossings in the extended (Sambe) space.
In particular, static detuning disorder drives eigenframe fragmentation near $\Omega_0=\omega_r$, while amplitude disorder drives fragmentation near $\Omega_0=2\omega_r$, where $\omega_r$ is the modulation frequency.
These windows are not merely regions of long coherence, but regimes in which the ensemble possesses a shared Floquet geometry.

We test these predictions experimentally in a nitrogen-vacancy (NV) ensemble in diamond.
We first realize long-lived collective population oscillations in the synchronized regime, then map robustness against controlled detuning and amplitude errors.
By sweeping the engineered gap at two modulation frequencies, we observe the predicted breakdown features near $\Omega_0=\omega_r$ and $\Omega_0=2\omega_r$, and verify that they move with $\omega_r$ rather than with a fixed laboratory frequency.
Finally, we use the synchronized regime to implement a continuous AC-sensing protocol whose response is a single tone at the modulation frequency, rather than a harmonic comb.
The same eigenframe diagnostic applies without modification beyond this particular adiabatic-following construction, as we illustrate for continuous and pulsed decoupling protocols in the Supplemental Material.
Together, these results identify the Floquet eigenframe itself as a measurable and engineerable resource for disorder-resilient ensemble control.

\section{\label{sec:AFS}Theory of Floquet Eigenframe Synchronization}

We develop the AFS theoretical framework in five stages. We first define the Floquet eigenframe of a driven ensemble and introduce the synchronization metrics that quantify its alignment across disorder realizations. We then identify the operational signatures that a synchronized ensemble exhibits. Next, we construct the modulated AFS drive that enforces Floquet adiabatic following at finite rate. We then introduce disorder and derive off-resonance criteria whose
unified interpretation is avoided crossings in the extended
(Sambe) space of the Floquet problem~\cite{shirley1965solution,
sambe1973steady}. Finally, we assemble these ingredients into a synchronization map that can be used to inform the experiment.

\subsection{Floquet eigenframes and synchronization metrics}

For an ensemble member $i$ evolving under a time-periodic Hamiltonian with period $T$, the Floquet propagator $U_i(T)$ has eigenvectors $\ket{F_\pm^i}$ that define the member's Floquet eigenframe. These definitions reference only the one-period propagator; the metrics below therefore apply to any periodically driven ensemble, continuous or pulsed, independent of the protocol that generates the drive, and we illustrate this protocol independence on concatenated continuous driving and pulsed CPMG in the Supplemental Material. We quantify the degree of synchronization across the ensemble through the order parameter
\begin{equation}
S_{\mathrm{AFS}} = \left| \frac{1}{N} \sum_i \hat n_i \right|,
\label{eq:SAFS}
\end{equation}
where $\hat n_i$ is the Bloch vector associated with $\ket{F_+^i}$ and $N$ is the total number of ensemble members. Here $\ket{F_+^i}$ is identified by adiabatic continuation from the disorder-free Floquet eigenstate (Supplemental Material). In the synchronized limit all ensemble members share a common Floquet quantization axis and $S_{\mathrm{AFS}} \to 1$; in the fragmented limit the Floquet eigenframes point in different directions and $S_{\mathrm{AFS}} \to 0$.

While $S_{\mathrm{AFS}}$ quantifies the collective alignment of Floquet eigenframes, experimental observables such as ensemble contrast are often sensitive to the spread of eigenframe orientations across the ensemble. We therefore introduce the complementary eigenframe fragmentation measure
\begin{equation}
\sigma_{\mathrm{AFS}} = \sqrt{1 - S_{\mathrm{AFS}}^2},
\label{eq:sigmaAFS}
\end{equation}
which increases monotonically with eigenframe spread, vanishing in the synchronized regime and approaching unity under complete fragmentation. For comparison with the disorder-free Floquet state we also define the overlap measure
\begin{equation}
Q_{\mathrm{AFS}} = \frac{1}{N} \sum_i \left| \langle F_+^{(0)} | F_+^i \rangle \right|^2,
\label{eq:QAFS}
\end{equation}
where $\ket{F_+^{(0)}}$ is the corresponding disorder-free Floquet eigenstate. This quantity measures average fidelity to the ideal Floquet eigenstate, but does not directly measure synchronization between ensemble members.

\subsection{\label{sec:signatures}Operational signatures of the synchronized regime}

In the synchronized limit, the one-period propagators of the ensemble share a single Bloch sphere rotation axis and differ primarily in rotation angle:
\begin{equation}
U_i(T) \simeq \exp\!\left[-\tfrac{i}{2}\,\theta_i\, \hat n\cdot\vec{\sigma}\right],
\label{eq:fibered}
\end{equation}
up to a member-dependent global phase. Here $\hat n$ is the common Floquet quantization axis and $\theta_i = \varepsilon_i T$ is set by the member's quasi-energy splitting. The approximation becomes exact for $S_{\mathrm{AFS}}=1$, while corrections are controlled by the eigenframe spread quantified by $\sigma_{\mathrm{AFS}}$ (Supplemental Material). This structure produces three linked operational signatures. First, the ensemble has a common Floquet quantization axis, measured directly by $S_{\mathrm{AFS}}$. Second, the population response factorizes: the geometric part of the stroboscopic trajectory is common across the ensemble, while residual ensemble dephasing enters through the distribution of quasi-energy phases (experimentally verified in Secs.~\ref{sec:Expts} and~\ref{sec:Robustness}). Third, weak signal perturbations transform into the Floquet frame through the same geometry for every member, giving a common filter-function structure with residual broadening set by quasi-energy dispersion (exploited in Sec.~\ref{ACMag}; Supplemental Material). When the eigenframes fragment, $\hat n$ becomes member-dependent, and initialization, readout, population dynamics, and signal response are averaged over inequivalent dressed bases. The synchronization metrics of Eqs.~\eqref{eq:SAFS}--\eqref{eq:sigmaAFS} therefore diagnose not merely a geometric alignment but the availability of collective control, collective coherence, and collective sensing.

\subsection{The AFS drive: enforcing Floquet adiabatic following}

Consider a two-level system driven by a time-dependent control field. In the rotating frame of the drive, the Hamiltonian is
\begin{equation}
H_0(t) = \frac{\hbar}{2} \left[ \Omega(t)\sigma_x + \Delta(t)\sigma_z \right],
\label{eq:H0}
\end{equation}
with sinusoidal modulation
\begin{equation}
\Omega(t) = \Omega_0 \sin(\omega_r t), \qquad \Delta(t) = -\Omega_0 \cos(\omega_r t),
\end{equation}
so that the effective field $\mathbf{b}(t) = (\Omega(t), 0, \Delta(t))$ has constant magnitude $\Omega_0$ and rotates in the $x$--$z$ plane at rate $\omega_r$. To enforce adiabatic following at finite modulation rate, we add the counterdiabatic (CD) correction
\begin{equation}
H_{\mathrm{cd}} = -\frac{\hbar}{2}\Omega_{\mathrm{cd}}\sigma_y = -\frac{\hbar}{2}\omega_r\sigma_y,
\end{equation}
giving the total control Hamiltonian
\begin{equation}
H_{\mathrm{c}}(t) = H_0(t) + H_{\mathrm{cd}}(t).
\end{equation}
Experimentally, the term $\Omega(t)\sigma_x$ is realized through amplitude modulation, while $\Delta(t)\sigma_z$ is realized through frequency modulation of the drive.

Although $H_{\mathrm{c}}(t)$ is time-dependent, this time dependence is removed in the \emph{control frame} defined by $U_r(t) = \exp(+i\,\omega_r t\,\sigma_y/2)$---a frame co-rotating with the geodesic at the modulation rate $\omega_r$. In this frame the CD term cancels the rotation generator and the control Hamiltonian collapses to the static form
\begin{equation}
H'_{\mathrm{c}} = -\frac{\hbar\Omega_0}{2}\sigma_z,
\label{eq:Hsync}
\end{equation}
with a constant dressed-state energy gap $\Omega_0$ (Supplemental Material). The Floquet eigenstates of $H_{\mathrm{c}}(t)$---the eigenvectors of the one-period propagator $U(T,0)$ with $T = 2\pi/\omega_r$---coincide in this frame with the eigenstates of $H'_{\mathrm{c}}$, namely $\ket{\uparrow}$ and $\ket{\downarrow}$. A spin initialized in a single Floquet eigenstate therefore remains in that eigenstate, appearing in the laboratory frame as a coherent population oscillation at the engineered rate $\omega_r$. In the disorder-free limit this construction realizes the synchronized regime exactly: every member possesses the same Floquet eigenframe, aligned with the control-frame axis $\hat z$.

\subsection{\label{sec:channels}Disorder channels and off-resonance criteria}

Static disorder enters through two dominant channels in our system. Amplitude inhomogeneity is a fractional variation of the drive field amplitude $\epsilon$, leading to
\begin{equation}
\Omega(t), \Omega_{\mathrm{cd}} \rightarrow (1+\epsilon)\Omega(t), (1+\epsilon)\Omega_{\mathrm{cd}},
\end{equation}
while $\Delta(t)$ is unaffected because it is generated through frequency modulation. The driven Hamiltonian becomes
\begin{align}
H(t) &= (1+\epsilon) H_{\mathrm{c}}(t) + V(t), \\
V(t) &= \frac{\hbar}{2}\epsilon\,\Omega_0 \cos(\omega_r t)\,\sigma_z,
\end{align}
where $V(t)$ represents the mismatch produced by the asymmetric action of amplitude errors on the amplitude- and frequency-modulated channels. Static detuning $\delta$ (a member-dependent shift of the bare transition frequency from the drive carrier) enters the rotating frame as
\begin{equation}
H_\delta = \frac{\hbar}{2}\delta\,\sigma_z.
\end{equation}

Transforming to the control frame makes the frequency content of each channel explicit. The detuning perturbation becomes
\begin{equation}
H'_\delta(t) = \frac{\hbar\delta}{2}\left[\cos(\omega_r t)\,\sigma_z + \sin(\omega_r t)\,\sigma_x\right],
\label{eq:Hdelta_sync}
\end{equation}
while the amplitude perturbation becomes
\begin{equation}
\begin{aligned}
H'_\epsilon(t) ={}& -\frac{\hbar\epsilon\Omega_0}{4}\,\sigma_z \;-\; \frac{\hbar\epsilon\omega_r}{2}\,\sigma_y \\
&+ \frac{\hbar\epsilon\Omega_0}{4}\left[\cos(2\omega_r t)\,\sigma_z + \sin(2\omega_r t)\,\sigma_x\right]
\end{aligned}
\label{eq:Heps_sync}
\end{equation}
(Supplemental Material). The diagonal ($\sigma_z$) components preserve the instantaneous control-frame eigenbasis and contribute only to Floquet phase evolution; in particular, the static part of Eq.~\eqref{eq:Heps_sync} is the channel through which the dressed gap $(1+\epsilon)\Omega_0$ disperses across the ensemble. The oscillatory $\sigma_x$ components are the resonant eigenvector-mixing channels responsible for finite-frequency eigenframe fragmentation: they occur at $\omega_r$ for static detuning and at $2\omega_r$ for amplitude inhomogeneity. The static $\sigma_y$ term in Eq.~\eqref{eq:Heps_sync} produces a small nonresonant tilt of the eigenframe from miscalibration of the CD correction. Each disorder source is thereby operator-resolved: tagged by the frequency of its resonant mixing component, and hence by the Floquet resonance at which it becomes significant.

These oscillatory eigenvector-mixing terms drive transitions across the engineered Floquet gap $\Omega_0$ only weakly when they remain off-resonant from it. In this regime, Floquet eigenframe synchronization is preserved. Inter-branch transitions are then dynamically suppressed and the ensemble retains a common Floquet eigenframe ($S_{\mathrm{AFS}} \approx 1$, $\sigma_{\mathrm{AFS}} \approx 0$), provided
\begin{equation}
\epsilon\,\Omega_0 \ll 8\,\abs{2\omega_r - \Omega_0}
\label{eq:crit_eps}
\end{equation}
for amplitude inhomogeneity and
\begin{equation}
\delta \ll 4\,\abs{\omega_r - \Omega_0}
\label{eq:crit_delta}
\end{equation}
for static detuning, where the numerical prefactors follow from a rotating-wave treatment of the mixing channels of Eqs.~\eqref{eq:Hdelta_sync} and~\eqref{eq:Heps_sync} (Supplemental Material). Each criterion expresses the same physical requirement: the disorder-induced eigenvector-mixing channel must remain sufficiently detuned from the Floquet gap to prevent fragmentation of the synchronized eigenframe. In the extended (Sambe) space of the Floquet problem, the two criteria acquire a unified interpretation: they mark avoided crossings between dressed Floquet branches, with the detuning channel exchanging one quantum of the modulation and becoming resonant at $\Omega_0 = \omega_r$, and the amplitude channel exchanging two quanta and becoming resonant at $\Omega_0 = 2\omega_r$ (Supplemental Material). Higher-order multi-photon resonances at other commensurabilities exist in principle, but their strengths are suppressed by additional powers of the disorder parameters and they lie outside the parameter range explored here (Supplemental Material).

\subsection{The synchronization map}

We now assemble these elements into a map of the synchronized regime. The synchronization criteria [Eqs.~\eqref{eq:crit_eps} and~\eqref{eq:crit_delta}] fail at three low-gap boundaries: the gap-closing point $\Omega_0\rightarrow0$, where no disorder channel is suppressed; the detuning-induced Floquet resonance $\Omega_0=\omega_r$, where the detuning eigenvector-mixing component becomes resonant with the Floquet gap; and the amplitude-induced Floquet resonance $\Omega_0=2\omega_r$, where the amplitude eigenvector-mixing component becomes resonant. Between and beyond these low-order resonances, off-resonant synchronization windows open. In the low-gap range, the first two are $0<\Omega_0<\omega_r$ and $\omega_r<\Omega_0<2\omega_r$; synchronization can also recover above $2\omega_r$ until higher-order resonances or other nonidealities become relevant. In these windows the Floquet eigenframes of different ensemble members remain aligned, corresponding to $S_{\mathrm{AFS}}\approx1$ and $\sigma_{\mathrm{AFS}}\approx0$: the ensemble retains a common Floquet quantization axis and evolves in a common Floquet geometry despite static disorder. As anticipated in Sec.~\ref{sec:signatures}, these are not merely regions of long coherence, but regimes in which collective control, coherence, and sensing remain available.

\begin{figure*}[t]
    \centering
    \includegraphics[width=\textwidth]{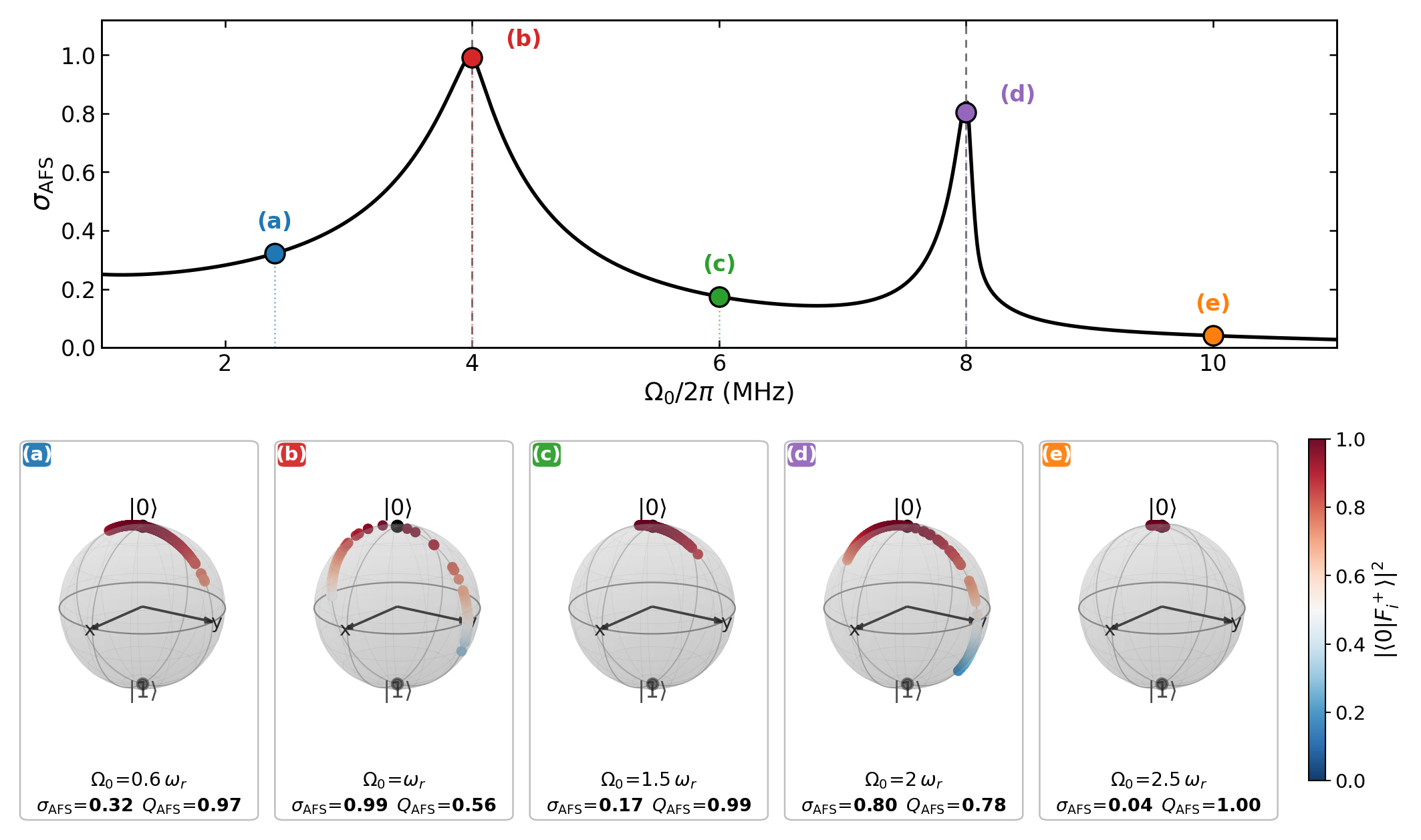}
    \caption{\textbf{Floquet eigenframe synchronization phase diagram.}
    (top)~Eigenframe fragmentation metric $\sigma_{\mathrm{AFS}}
    = \sqrt{1 - S_{\mathrm{AFS}}^2}$ versus engineered gap $\Omega_0$ at
    fixed modulation rate $\omega_r$, computed for an ensemble with
    amplitude disorder $\sigma_\epsilon$ = 1\% and effective detuning disorder
    $\sigma_\delta$ = 0.9 MHz. Dotted vertical lines mark the predicted
    synchronization breakdowns at $\Omega_0=\omega_r$ (detuning channel)
    and $\Omega_0=2\omega_r$ (amplitude channel).
    (bottom)~Floquet-eigenvector Bloch-sphere distributions sampled at
    $\Omega_0/\omega_r = 0.6,\,1,\,1.5,\,2,\,2.5$ as indicated in the (top) plot and colored by overlap
    $|\langle F_+^{(0)}|F_+^i\rangle|^2$. Synchronized regions show tight
    clustering near a common Floquet quantization axis; resonance
    breakdowns scatter the eigenframes across the Bloch sphere.}
    \label{fig:afs_phase_diagram}
\end{figure*}

The resulting structure is illustrated in Fig.~\ref{fig:afs_phase_diagram}, which shows the evolution of the eigenframe fragmentation measure $\sigma_{\mathrm{AFS}}$ together with representative Floquet eigenframe distributions across the ensemble. Away from the Floquet resonances, the Floquet eigenvectors remain tightly clustered around a common quantization axis, indicating synchronized evolution. As the resonance conditions $\Omega_0=\omega_r$ and $\Omega_0=2\omega_r$ are approached, disorder drives resonant mixing between Floquet branches, causing the eigenframes to rotate in member-dependent directions and fragment across the Bloch sphere. This fragmentation corresponds to large values of $\sigma_{\mathrm{AFS}}$ and marks the breakdown of synchronization.

The synchronization map provides a direct visualization of the central phenomenon studied in this work: the emergence, persistence, and eventual breakdown of Floquet eigenframe synchronization in a disordered periodically driven ensemble. In Sec.~\ref{sec:criterion}, we test these predictions experimentally by sweeping $\Omega_0$ and locating the predicted synchronization breakdown points.

\begin{figure}[h]
  \centering
  \includegraphics[width=\linewidth]{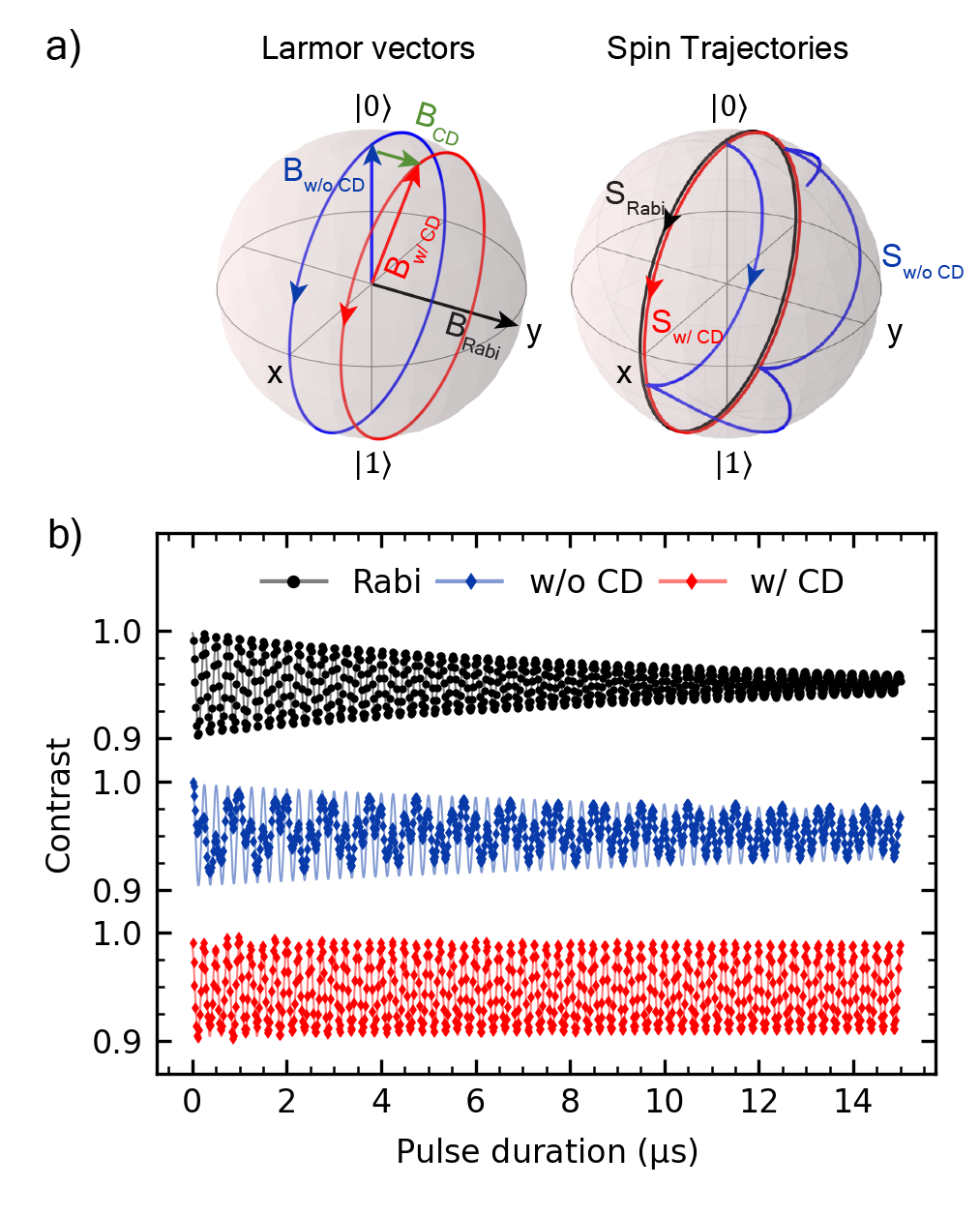}  
  \caption{\textbf{Robust ensemble population control via AFS.}
(a) Bloch-sphere visualization of the effective control fields and spin trajectories for CW Rabi driving (black), modulated driving without counterdiabatic (CD) correction (blue), and full AFS protocol with CD correction (red). The CD term enforces Floquet adiabatic following at the modulation rate used here $\omega_r/2\pi = 4$~MHz.
(b) Measured NV ensemble photoluminescence (PL)   oscillations for the three protocols, resulting from coherent population modulation. The CW Rabi oscillation decays within $\approx 5~\mu$s due to ensemble inhomogeneity, while the full AFS protocol preserves high-contrast oscillations beyond $15~\mu$s, demonstrating robust collective population control.}
  \label{fig:1}
\end{figure}

\section{\label{sec:Expts}Experimentally Realizing the Synchronized Regime: Collective Rabi Oscillations}

Coherent population (i.e. Rabi) oscillations provide the natural first experimental probe of the synchronized regime. In a conventional continuous-wave (CW) Rabi experiment on an inhomogeneous ensemble, static disorder degrades the collective oscillation through the two effects emphasized in Sec.~\ref{sec:AFS}: tilting the rotation axes of different members, fragmenting their Floquet eigenframes; and dispersing member rotation rates through quasi-energy disorder. The latter effect is expected to be especially visible in CW driving because the initial state $\ket{0}$ is an equal superposition of the drive's dressed eigenstates, so the population signal contains an interference term between Floquet branches that samples the full quasi-energy spread. The ensemble-averaged signal therefore dephases rapidly.

The AFS drive suppresses both effects. In the synchronized regime the eigenframes align across the ensemble; and because the disorder-free AFS Floquet eigenstate coincides with $\ket{0}$ at stroboscopic times (Supplemental Material), the prepared state has near-unity overlap with the same Floquet branch for all ensemble members. The inter-branch interference term is therefore suppressed, and residual quasi-energy dispersion couples only weakly to the population response. Long-lived collective oscillations are thus the expected direct experimental effect of the factorized response described in Sec.~\ref{sec:signatures}.

To experimentally realize this behavior, we apply a frequency- and amplitude-modulated microwave field to an ensemble of NV centers in diamond addressed with a confocal microscope ($^{15}$N-enriched sample and setup details in the Supplemental Material). The NV ground-state Hamiltonian in the laboratory frame is
\begin{equation}
\begin{aligned}
\frac{H_{\mathrm{NV}}}{\hbar} =\ & D S_z^2 \;-\; \gamma_e\,\vec{B}\cdot\vec{S}
   \;+\; \vec{S}\cdot\mathbf{A}\cdot\vec{I} \\
   &+ \sqrt{2}\,\Omega(t)\cos\!\left(\omega_{\mathrm{MW}}\,t
        - \int_0^{t}\!\Delta(t')\,dt' + \phi\right) S_x ,
\end{aligned}
\label{eq:HNV}
\end{equation}
where $D/2\pi \approx 2.87$~GHz is the zero-field splitting,
$\gamma_e/2\pi \approx 2.8024$~MHz/G is the NV gyromagnetic
ratio~\cite{doherty2013,barry2020}, $\vec{S}$ ($\vec{I}$) is the electron spin-$1$
($^{15}$N nuclear spin-$1/2$) operator vector, $\mathbf{A}$ is the diagonal
hyperfine tensor, $\omega_{\mathrm{MW}}$ is the microwave carrier
frequency, and $\Omega(t)$ and $\Delta(t)$ are the time-dependent transverse
drive amplitude and detuning, respectively. The factor $\sqrt{2}$ absorbs the spin-$1$
matrix element $\langle 0|S_x|\!\pm\!1\rangle = 1/\sqrt{2}$, so $\Omega(t)$
is the Rabi frequency within the driven two-level subspace. Tuning
$\omega_{\mathrm{MW}}$ onto resonance with one of the NV ground-state
transitions, e.g.,\ $\ket{0}\leftrightarrow\ket{-1}$, isolates an effective
two-level subsystem.

In the rotating frame defined by $\omega_{\mathrm{MW}}$ via
$U_{\mathrm{MW}}(t) = \exp\!\left[-i\,(\omega_{\mathrm{MW}}\,t/2)\,\sigma_z\right]$, the
Hamiltonian reduces to Eq.~\eqref{eq:H0}, allowing direct implementation of
the disorder control techniques developed in Sec.~\ref{sec:AFS}. The control fields are
synthesized by an arbitrary waveform generator (Keysight M8190A), which
enables simultaneous amplitude and frequency modulation. This realizes a
control field trajectory defined by $\Omega(t)$ and $\Delta(t)$, with and without
the CD correction.

All measurements reported in this section are performed at an engineered modulation rate
$\omega_r/2\pi = 4$~MHz with dressed gap $\Omega_0/2\pi = 3$~MHz, placing
the experiment in the first synchronization map window,
$0 < \Omega_0 < \omega_r$. Figure~\ref{fig:1}(a) shows Bloch-sphere visualizations of the three control
protocols employed in these measurements: (i) conventional CW Rabi driving, (ii) modulated driving
without the CD correction, and (iii) full AFS driving with the CD correction. The
corresponding measured NV ensemble photoluminescence (PL) as a function of drive pulse duration is
shown in Fig.~\ref{fig:1}(b). Under CW Rabi driving, the ensemble-averaged oscillating PL
signal loses contrast within $\approx 5~\mu$s, consistent with eigenframe
fragmentation and quasi-energy dispersion acting unchecked. The modulated drive without the
CD correction improves stability but still decays, with
residual inter-branch coupling modulating the signal. The full AFS drive
instead produces high-contrast PL oscillations that persist over the full
measurement window.

These long-lived PL oscillations are a clear experimental signature of the synchronized regime, consistent with the theoretical framework developed above. At the chosen experimental operating point, both off-resonance criteria of Eqs.~\eqref{eq:crit_eps}--\eqref{eq:crit_delta} are comfortably satisfied, so the NV ensemble evolves within a common Floquet eigenframe, and the AFS initial condition projects predominantly onto a single Floquet branch, as required for the factorized response of Sec.~\ref{sec:signatures}. The static disorder that rapidly dephases conventional CW Rabi driving therefore neither appreciably fragments the eigenframes nor couples strongly to the measured population observable. Further experiments described below test how far this protection extends along the disorder axes (Sec.~\ref{sec:Robustness}), and map where it fails (Sec.~\ref{sec:criterion}).

\section{\label{sec:Robustness}Disorder Robustness in the Synchronized Regime}
\begin{figure}[h]
  \centering
  \includegraphics[width=\linewidth]{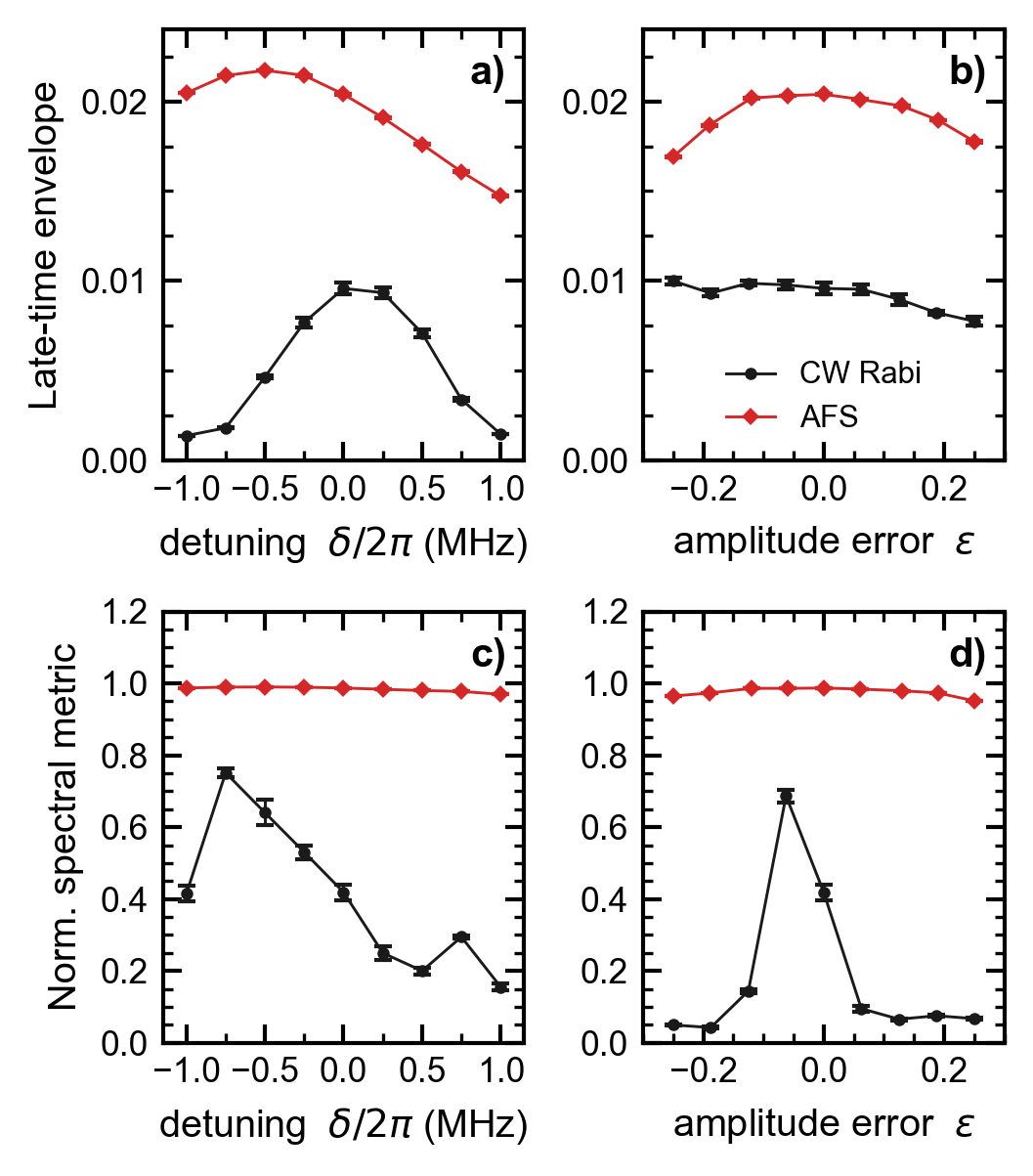}  
  \caption{\textbf{Robustness of the synchronized ensemble dynamics.} Measured late-time oscillation metrics under controlled detuning and amplitude miscalibration, comparing CW Rabi driving (black) with the AFS protocol (red) at the operating point $\omega_r/2\pi=4$~MHz, $\Omega_0/2\pi=3$~MHz. The panels are line cuts of a two-dimensional $(\epsilon,\delta)$ scan, through $\epsilon=0$ [(a),(c)] and $\delta=0$ [(b),(d)]. (a,b)~Mean Hilbert-envelope amplitude over the $5$--$10~\mu$s late-time window, versus detuning $\delta$ and fractional amplitude error $\epsilon$ respectively. (c,d)~Normalized spectral metric---the Fourier amplitude at $\omega_r$ divided by the mean Hilbert-envelope amplitude over the same window---for the same two cuts. AFS sustains both contrast and frequency lock across the full range, whereas CW Rabi driving loses contrast under detuning and frequency fidelity under amplitude error. Error bars denote the standard error of the mean over five measurement repetitions.}
  \label{fig:robustness}
\end{figure}

Having experimentally realized the synchronized regime, we now test one of its central predictions: Floquet eigenframe synchronization should persist under substantial amplitude and detuning disorder, provided the synchronization criteria of Eqs.~\eqref{eq:crit_eps} and~\eqref{eq:crit_delta} remain satisfied. To perform this experimental test, we keep the operating point fixed at $\omega_r/2\pi = 4$~MHz and $\Omega_0/2\pi = 3$~MHz, as in the previous section, and systematically vary the ensemble-averaged detuning $\delta$ and fractional amplitude error $\epsilon$ while keeping their intrinsic distributions unchanged. We then compare the coherence of the AFS drive and conventional CW Rabi driving across the resulting two-dimensional $(\epsilon,\delta)$ parameter space. From each measured late-time PL trace (i.e., $5$--$10~\mu$s window) we extract two complementary metrics: (i) the mean Hilbert-envelope amplitude, which tracks surviving collective contrast; and (ii) the normalized spectral metric---the Fourier amplitude at $\omega_r$ divided by the mean Hilbert-envelope amplitude over the same window---which characterizes whether the ensemble still oscillates at the drive-imposed frequency. The second metric is a direct probe of the factorized response of Sec.~\ref{sec:signatures}: in the synchronized regime, the oscillation frequency is set by the modulation rate $\omega_r$, common to every ensemble member, whereas under CW driving it is set by the member-dependent generalized Rabi rate. Figure~\ref{fig:robustness} summarizes the experimental results as  line cuts through the two-dimensional ($\epsilon, \delta$) parameter space at $\epsilon=0$ and $\delta=0$; error bars denote the standard error over five experimental repetitions.

Along the detuning axis [Figs.~\ref{fig:robustness}(a,c)], CW Rabi driving loses contrast rapidly: its late-time envelope amplitude falls to near the noise floor by $\delta/2\pi = \pm1$~MHz as static inhomogeneity dephases the NV ensemble. The AFS protocol instead sustains a large, slowly varying envelope amplitude across the full range: roughly twice the on-resonance value for CW Rabi driving, and an order of magnitude above CW Rabi at the extremes. This result is consistent with the synchronization criterion prediction: Eq.~\eqref{eq:crit_delta} bounds robust operation by $\delta \ll 4\,\abs{\omega_r-\Omega_0}$, which at this experimental operating point is $\delta/2\pi \ll 4$~MHz, so the $\pm1$~MHz scan lies within the synchronized regime. The mild decline of the AFS envelope amplitude toward positive detuning reflects the growing ratio $\delta/\Omega_0$ and the approach to a neighboring $^{15}$N hyperfine transition, for which the counterdiabatic (CD) correction is not optimized. The normalized spectral metric [Fig.~\ref{fig:robustness}(c)] sharpens the distinction: the AFS oscillation stays locked at $\omega_r$ across the entire range, while the CW Rabi spectral weight at $\omega_r$ is suppressed and irregular, sitting well below the AFS value even at zero offset, where residual intrinsic inhomogeneity already broadens the CW Rabi response. Together, these observations indicate that the Floquet eigenframe remains synchronized despite substantial detuning offsets.

Measurements along the amplitude axis [Figs.~\ref{fig:robustness}(b,d)] show similar behavior. The AFS envelope varies only weakly over the scanned range $\epsilon = \pm25\%$, and its spectral metric remains pinned near unity throughout; CW Rabi driving retains comparable contrast but preserves spectral weight at $\omega_r$ only within a narrow window about $\epsilon=0$, since the bare Rabi frequency tracks the drive amplitude directly and walks away from $\omega_r$ as $\epsilon$ grows, while the AFS oscillation remains pinned at the drive-imposed rate. These results also follow the synchronization criterion: Eq.~\eqref{eq:crit_eps} gives $\epsilon\,\Omega_0 \ll 8\,\abs{2\omega_r-\Omega_0}$, an amplitude tolerance far wider than the present experimental scan and consistent with the observed near-flat AFS response. The persistence of both contrast and spectral purity therefore indicates that Floquet eigenframe synchronization remains intact across the full amplitude scan.

Together, these experimental results map the synchronized regime along the disorder axes $\epsilon$ and $\delta$ at fixed $\Omega_0$, confirming that it is broad but, as Section~\ref{sec:AFS} anticipates, bounded. The complementary cut through parameter space---holding the disorder fixed and instead varying $\Omega_0$, so that the off-resonance offsets $\abs{\omega_r-\Omega_0}$ and $\abs{2\omega_r-\Omega_0}$ are themselves driven toward zero---maps that boundary directly, and is the subject of Section~\ref{sec:criterion}.

\section{\label{sec:criterion}Mapping the Synchronization Breakdown Spectrum}

The framework of Sec.~\ref{sec:AFS} makes its sharpest prediction here: synchronization must fail whenever a disorder-induced eigenvector-mixing channel becomes resonant with the engineered Floquet gap: at $\Omega_0 = \omega_r$ through the detuning channel of Eq.~\eqref{eq:Hdelta_sync}, and at $\Omega_0 = 2\omega_r$ through the amplitude channel of Eq.~\eqref{eq:Heps_sync}. Between these resonances the ensemble should remain synchronized. Crucially, the predicted breakdown positions contain no adjustable parameter: they are set by the commensurability ratio $\Omega_0/\omega_r$ alone, and must therefore move with the modulation rate. We experimentally test this prediction directly by sweeping $\Omega_0$ while holding $\omega_r$ fixed. At each $\Omega_0$ we record a time-resolved PL contrast trace under the full AFS drive and extract its late-time oscillation envelope, the mean Hilbert-envelope amplitude over the $5$--$10~\mu$s window; note that a synchronized operating point sustains a large-amplitude late-time envelope, while a breakdown point collapses it. To separate breakdown governed by the ratio $\Omega_0/\omega_r$ from breakdown set by any fixed frequency laboratory systematic effect, we perform the measurements at two modulation rates, $\omega_r/2\pi = 2$ and $3$~MHz.

\begin{figure}[t]
  \centering
  \includegraphics[width=\linewidth]{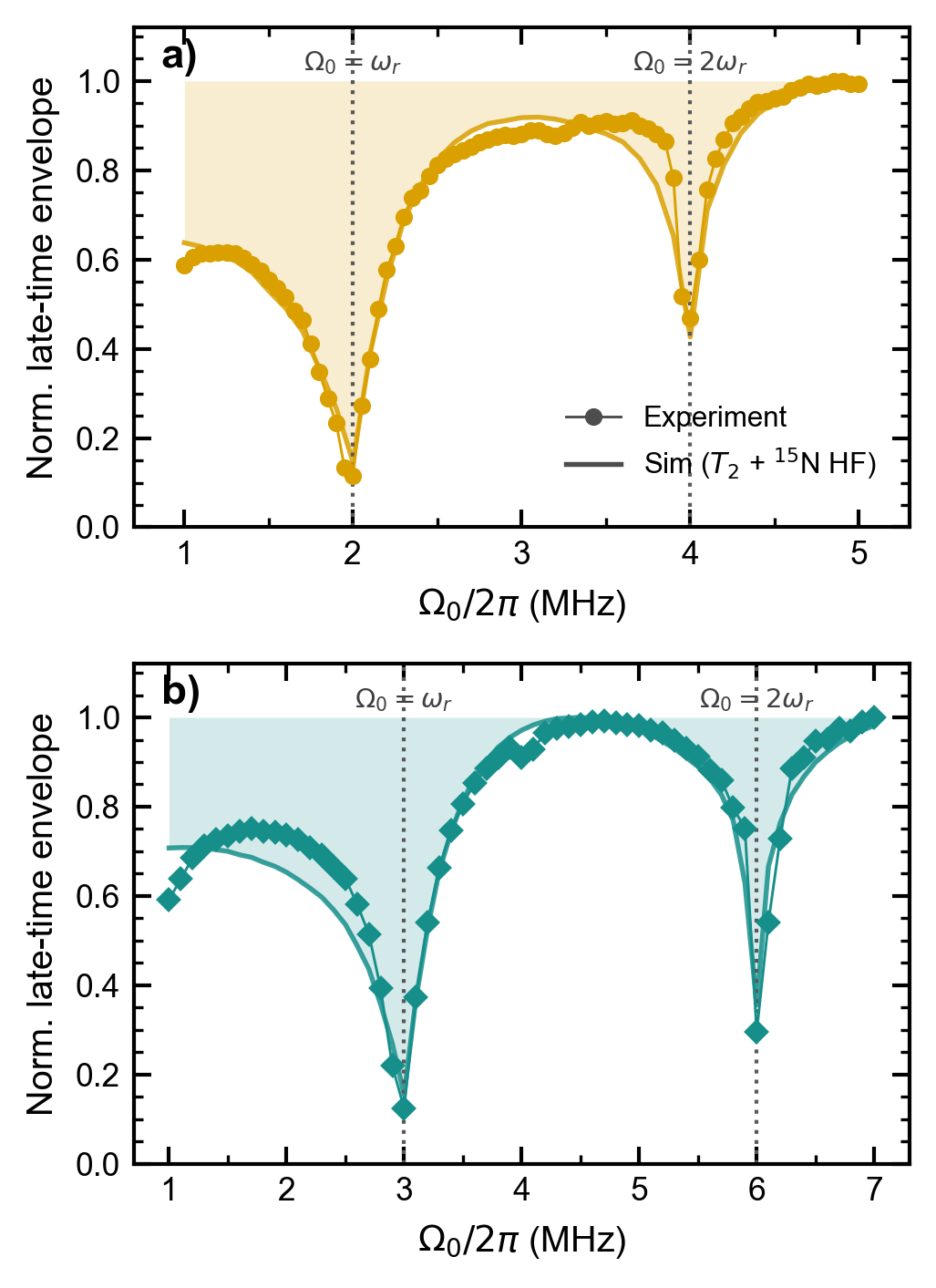}
  \caption{\textbf{Mapping the synchronization breakdown spectrum.} Normalized late-time oscillation envelope (mean Hilbert-envelope amplitude of the AC component over the $5$--$10~\mu$s window of the measured PL time trace) as the engineered gap $\Omega_0$ is swept at fixed modulation rate, for (a) $\omega_r/2\pi = 2$~MHz and (b) $\omega_r/2\pi = 3$~MHz. Markers are experiment; the solid curve in each plot is a matched-distribution simulation of an ensemble with amplitude disorder $\sigma_\epsilon = 1\%$ and effective detuning disorder $\sigma_\delta/2\pi = 0.9$~MHz (Supplemental Material). Each simulation curve is normalized to its own maximum; and the shaded region marks the envelope amplitude deficit relative to the synchronized baseline. Dotted lines mark the predicted breakdown points $\Omega_0 = \omega_r$ and $\Omega_0 = 2\omega_r$. Both the measured and simulated envelope amplitudes collapse sharply at both points, and the dips shift with $\omega_r$ between the two runs, confirming that breakdown is governed by the ratio $\Omega_0/\omega_r$.}
  \label{fig:criterion}
\end{figure}

Figure~\ref{fig:criterion} shows the results. For both modulation rates, the late-time envelope collapses sharply at two values of $\Omega_0$, coinciding with the predicted breakdown points $\Omega_0 = \omega_r$ and $\Omega_0 = 2\omega_r$ to within the sweep step. When $\omega_r/2\pi$ is changed from $2$ to $3$~MHz the dips move with it---from $\Omega_0/2\pi = 2$ and $4$~MHz to $3$ and $6$~MHz---confirming that breakdown is governed by the commensurability ratio rather than any fixed frequency scale. Between the two dips the envelope amplitude recovers close to its maximum, marking the robust window $\omega_r < \Omega_0 < 2\omega_r$ in which both off-resonance criteria are comfortably satisfied. Toward small $\Omega_0$ the envelope declines once more, indicative of the third breakdown anticipated in Sec.~\ref{sec:channels}: as $\Omega_0\to0$ the modulated $\sigma_x$--$\sigma_z$ control field vanishes, leaving only the counterdiabatic (CD) $\omega_r\sigma_y$ term; the Floquet gap that protects synchronization closes, and the protocol loses the synchronized-frame structure that distinguishes it from ordinary continuous driving.

We perform a matched-distribution simulation that propagates an ensemble with amplitude disorder $\sigma_\epsilon = 1\%$ and effective detuning disorder $\sigma_\delta/2\pi = 0.9$~MHz (Supplemental Material), and includes static inhomogeneity, intrinsic relaxation, and the $^{15}$N hyperfine splitting. This simulation reproduces the measured structure across both experimental runs with a single common set of ensemble parameters: the observed dip positions, recovery window, and small-$\Omega_0$ rolloff. The two dips are moreover operator-resolved in the sense of Sec.~\ref{sec:channels}: i.e., the collapse at $\Omega_0 = \omega_r$ corresponds to the detuning channel and the collapse at $\Omega_0 = 2\omega_r$ to the amplitude channel, so the sweep does not merely locate the breakdowns, but also identifies which disorder operator causes each.

Taken together---the coincidence of the collapse points with the predicted resonance conditions, their scaling with $\omega_r$ across independent measurements, their localized form, and the quantitative agreement with matched-distribution simulation---these measurements provide direct experimental evidence for Floquet eigenframe synchronization and its resonance-governed breakdown. They constitute the experimental realization of the synchronization map derived in Sec.~\ref{sec:AFS} and computed in Fig.~\ref{fig:afs_phase_diagram}.

\begin{figure*}[t]
    \centering
    \includegraphics[width=\textwidth]{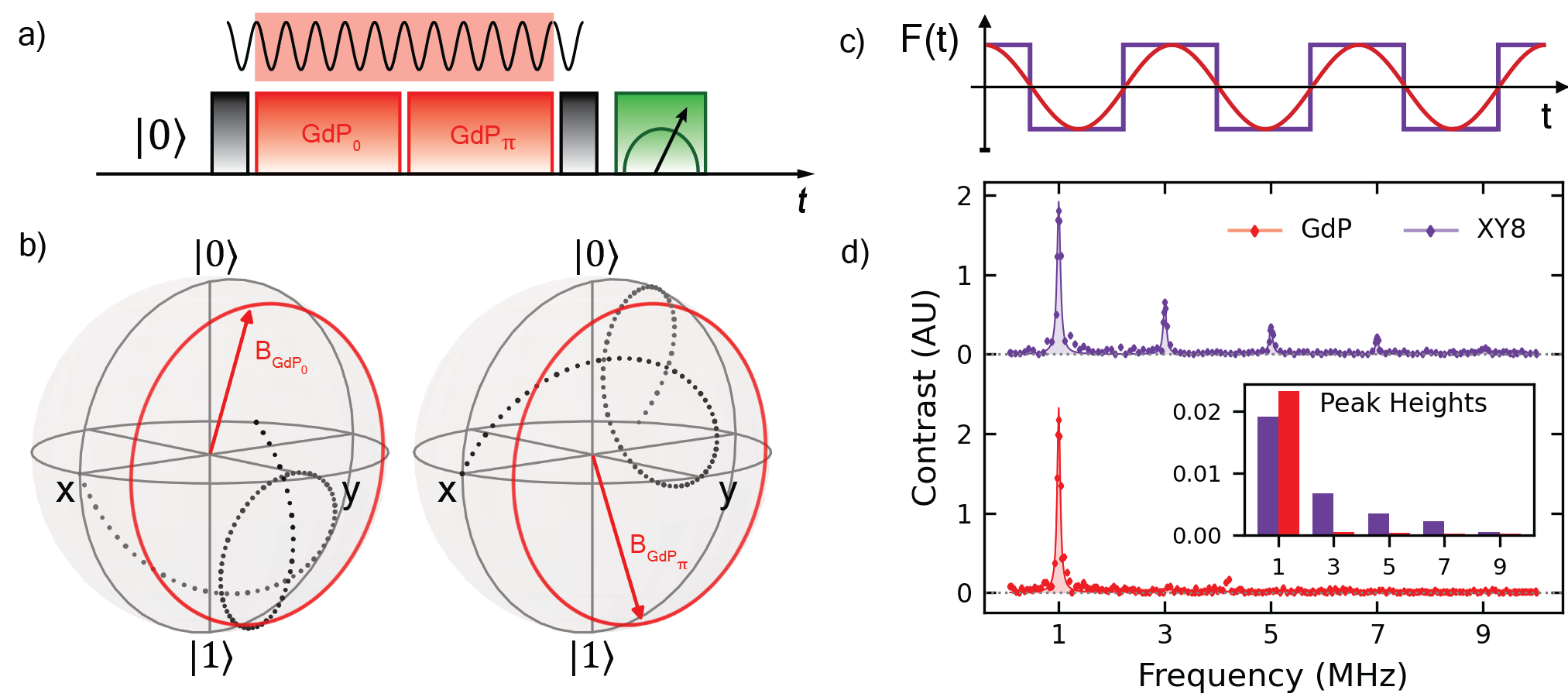}
    \caption{
    \textbf{Synchronization-enabled continuous AC sensing.}
    (a) Schematic of the continuous AFS sensing sequence $[\mathrm{GdP}_0, \mathrm{GdP}_\pi]$.
    (b) Bloch-sphere trajectories showing cancellation of the dynamic phase between the two geodesic blocks.
    (c) Filter functions for XY8 (purple) and the continuous AFS protocol (red).
    (d) Experimental AC magnetometry spectra, demonstrating the single-tone response of the synchronized NV ensemble alongside the odd-harmonic comb of XY8.
    }
    \label{fig:geodesic_protocol}
\end{figure*}

\section{\label{ACMag}Synchronization-Enabled Continuous AC Magnetometry}

Having experimentally verified Floquet eigenframe synchronization, we now demonstrate its operational utility for AC magnetometry using the experimental content of the common-filter signature of Sec.~\ref{sec:signatures}. In the synchronized regime, all NV ensemble members share a common Floquet quantization axis despite static disorder, so an external AC perturbation projects into the same Floquet eigenframe for every spin: the ensemble responds as a single collective sensor with a well-defined filter function. Without synchronization, each spin acquires a different sensing response and the collective spectral selectivity washes out in the ensemble average.

We employ an AC signal sensing sequence that builds on the geodesic-evolution filter design of Zeng \emph{et al.}~\cite{zeng2024}, but uses the synchronized Floquet eigenframe established above to make the single-tone response collective across an inhomogeneous ensemble.

The continuous-drive AC sensing protocol [Fig.~\ref{fig:geodesic_protocol}(a)] consists of: (i) a laboratory $\pi/2$ pulse preparing an equal superposition of dressed eigenstates; (ii) a geodesic block of the AFS drive, $\mathrm{GdP}_0$, of duration $n_p(2\pi/\omega_r)$, during which the eigenframe traces a geodesic at rate $\omega_r$; (iii) a second block $\mathrm{GdP}_\pi$ of equal duration with the drive phase advanced by $\pi$; and (iv) a final $\pi/2$ pulse for NV ensemble PL readout. The integer-period constraint on each segment returns the Floquet eigenframe to a fixed orientation at every block boundary, so the dressed coherence is well defined at the handoff between segments. Unlike the population measurements of Sec.~\ref{sec:Expts}, the sensing sequence deliberately prepares a superposition of Floquet branches, so the quasi-energy dispersion $(1+\epsilon)\Omega_0$ that synchronization leaves unaffected now couples to the target AC signal measurement. The $\pi$ phase advance between the two blocks inverts the sign of the effective field seen by each NV spin in the ensemble; and the pair $[\mathrm{GdP}_0, \mathrm{GdP}_\pi]$ therefore acts as an echo within the synchronized Floquet frame, refocusing this residual eigenvalue dispersion at first order [Fig.~\ref{fig:geodesic_protocol}(b)]. This echo is possible precisely because it refocuses eigenvalue disorder about a shared axis, i.e., the refocusable kind of disorder identified in Sec.~\ref{sec:Intro}.

Within the integer-period window, an AC signal coupling to the NV sensor spins via  $\sigma_z$ transforms into the rotating Floquet frame as a sinusoidal modulation at $\omega_r$ along the common quantization axis (Supplemental Material). The passband of the resulting filter is set by the modulation waveform~\cite{kotler2011}; and because the effective modulation here is sinusoidal, the single-spin filter is a single passband centered at $\omega_r$. This contrasts with pulsed dynamical-decoupling sequences, whose square-wave toggling functions generate a comb of odd harmonics at $(2k+1)\omega_r$~\cite{alvarez2011,degen2017}, augmented under realistic pulse errors by spurious even harmonics and subharmonics~\cite{loretz2015}; and with phase-switched continuous schemes such as rotary echoes, which inherit analogous harmonic structure~\cite{aiello2013,louzon2025}. The single-tone response of the present approach is a property of the sinusoidal effective modulation and holds within the rotating-wave regime realized here; it is not a consequence of continuity alone. A concurrent theoretical proposal reaches an analogous single-peak filter with a purely adiabatic axis-steering drive on a single quantum probe or sensor~\cite{li2026lockin}: there the modulation rate is bounded by the adiabatic condition, whereas the CD term of the AFS drive considered here lifts this constraint. Also, the ensemble question does not arise for a single probe as treated in \cite{li2026lockin}. Floquet eigenframe synchronization plays a distinct and essential role in the present approach: it ensures that the single-tone filter is identical across the ensemble despite static disorder, so the spectral selectivity survives collective averaging, with residual broadening set by the quasi-energy dispersion that the $[\mathrm{GdP}_0,\mathrm{GdP}_\pi]$ echo suppresses. Figure~\ref{fig:geodesic_protocol}(c) compares the computed filter functions of the continuous AFS protocol and the discretized XY8 pulse sequence at a matched fundamental frequency. Figure~\ref{fig:geodesic_protocol}(d) shows the measured AC magnetometry spectra: the synchronized continuous protocol responds at the single tone $\omega_r$, while XY8 exhibits its characteristic odd-harmonic comb.

The continuous implementation also lifts a bandwidth constraint of discretized adiabatic protocols. In a discretized realization the accessible sensing frequency is bounded by $2(\omega_r)_\mathrm{max}/N$, where $N$ is the number of pulses required to approximate the geodesic trajectory, so increasing $\omega_r$ trades fidelity against bandwidth. In the continuous implementation, the bandwidth is set directly by the maximum achievable CD amplitude $\Omega_\mathrm{cd}=\omega_r$, with no pulse-count penalty.

Three roles separate cleanly in the continuous AFS protocol. (1) The sinusoidal geodesic modulation determines the filter shape, producing a harmonic-free single-tone response centered at $\omega_r$. (2) Floquet eigenframe synchronization establishes a common Floquet geometry across the ensemble, ensuring that this filter is shared by all spins and therefore survives collective averaging. (3) The $[\mathrm{GdP}_0,\mathrm{GdP}_{\pi}]$ echo suppresses the residual eigenvalue dispersion within the synchronized frame, extending the lifetime of the collective response. Together these ingredients produce an AC sensing protocol that is simultaneously spectrally selective, collective, and robust against static disorder. The synchronized-frame echo can also be repeated as a multi-block outer envelope, giving a natural route to CPMG-style coherence extension
under colored bath noise without changing the inner AFS geometry
(Supplemental Material).

\section{Discussion and Outlook}
\label{sec:discussion}

As developed here Floquet eigenframe synchronization exploits the interplay of three approaches to quantum control applied to ensembles:

\emph{Quasienergy protection.} Continuous dressing of an inhomogeneously broadened collective excitation has a clear precedent in atomic ensembles, where Finkelstein \emph{et al.}~\cite{finkelstein2021} suppressed Doppler dephasing by dressing the collective state with an auxiliary level of opposite susceptibility, cancelling the first-order eigenvalue shift. Concatenated continuous dynamical decoupling (CCDD)~\cite{cai2012, stark2017, wang2020} and its recent extension to GHz AC magnetometry using NV ensembles~\cite{kitamura2025} flatten a dressed quasienergy gap against amplitude and detuning noise by nesting a second drive in the rotating frame of the first. These strategies can be engineered and analyzed in the language of quasienergy protection. The synchronization framework established here supplies the complementary ensemble-level geometric variable: applying the order parameter to CCDD shows that its working regime coincides with eigenframe alignment across the ensemble (Supplemental Material) and thereby continuous dressing protects where it synchronizes. Within this class, what distinguishes adiabatic frame synchronization (AFS) is not eigenframe alignment as such, but the automatic satisfaction of a second, logically independent condition: the AFS Floquet eigenstate coincides with the bare initialization state, so the aligned frame is visible in the population observable without dressed-state preparation, and the $[\mathrm{GdP}_0, \mathrm{GdP}_\pi]$ echo then refocuses the residual quasienergy dispersion that synchronization leaves at first order.

\emph{Average-Hamiltonian engineering on disordered spin ensembles.} Pulsed
Hamiltonian-engineering protocols achieve disorder robustness via algebraic
cancellation in the Magnus expansion. Recent qudit Hamiltonian-engineering
work on the same (electronic spin-1) NV-ensemble platform~\cite{zhou2024prx} demonstrated $>10\times$ spin coherence improvement over prior pulse sequences using a graphical
$S_z$-transformation framework with explicit robustness conditions against
control errors and on-site disorder. Related Floquet studies in dense NV
ensembles have also shown that high-frequency periodic and quasiperiodic
driving can stabilize long-lived prethermal dynamics by suppressing energy
absorption~\cite{he2023}. These approaches protect driven
dynamics through average-Hamiltonian or prethermal effective-Hamiltonian
mechanisms. AFS instead aligns the ensemble within a common Floquet
eigenframe through a single continuous counterdiabatic (CD) drive, makes that
alignment directly measurable through the order parameter
$S_{\mathrm{AFS}}$, and identifies the low-order breakdown resonances at
$\Omega_0=\omega_r$ and $\Omega_0=2\omega_r$ as the first boundaries of
robust operation. Importantly, we find that these metrics are protocol-independent. The Supplemental Material applies them, unchanged, to pulsed CPMG on the same NV ensemble: $S_{\mathrm{AFS}}$ is found to decay with pulse spacing in quantitative agreement with a closed-form prediction; and the eigenframe fragmentation correlates with the observed ensemble decoherence. Equally important is what the diagnostic metrics do \emph{not} capture: as shown in the Supplemental Material, at large pulse spacing CPMG retains a residual envelope amplitude plateau from its echo mechanism even though the eigenframes are fully fragmented. The AFS approach therefore identifies synchronization-based protection precisely where it operates, without subsuming other protection mechanisms such as echo cancellation or average-Hamiltonian suppression, a scope we consider a feature: it makes the diagnostic falsifiable and mechanism-specific rather than a proxy for generic coherence.

\emph{Counterdiabatic Floquet control.} The variational CD primitive~\cite{claeys2019}, recently extended to periodically driven systems~\cite{schindler2024, schindler2025}, guarantees transitionless state transfer between Floquet eigenstates of a single Hamiltonian. AFS exploits this primitive at the ensemble level, where the protected
object becomes the cross-realization Floquet frame and the figure of
merit becomes the Kuramoto-style alignment order parameter
$S_{\rm AFS}$~\cite{acebron2005kuramoto} rather than single-state
fidelity. We expect the same construction to extend to other periodically driven inhomogeneous quantum ensembles: e.g., atomic clouds with Doppler or Zeeman spread, trapped-ion crystals with mode-frequency dispersion, superconducting qubit arrays with charge or flux disorder, and solid-state donor or color-center systems; i.e., wherever a control geometry supports a CD correction.

A related breakdown of disorder-protected Floquet dynamics has recently been reported in a hyperpolarized $^{13}$C nuclear spin dipolar network in diamond, where stochastic electron-spin switching tunes rare $^{13}$C clusters into multi-photon resonance~\cite{selco2026}. In this case the driven spins are nuclei, the disorder is dynamical, the observable is a prethermal heating rate, and the protected object is a disorder-stabilized prethermal plateau. In the present case, the driven spins are electronic, the disorder is static, the observable is an explicit ensemble eigenframe-synchronization order parameter, and the protected object is the Floquet frame itself. The shared theme---disorder-induced Floquet structure breaking at sharp parametric resonances---underscores the broader relevance of the resonant ordering principle introduced here. Concurrent NV-ensemble Floquet spectroscopy work~\cite{nguyen2026}
analyzes finite-pulse WAHUHA control through the detuning-dependent
Floquet eigenphase, showing that phase wrapping and quasi-energy branch
folding can extend the stroboscopic decay time while suppressing the
detuning-to-phase transduction slope relevant for DC sensing. The present
work addresses the complementary eigenvector problem: whether the
one-period propagators of different disorder realizations share a common
Floquet eigenframe, and where disorder-induced mixing fragments that
frame.

More broadly, the present results identify the Floquet \emph{eigenframe} itself, not only the Floquet \emph{spectrum}, as a measurable and engineerable resource for periodically driven disordered ensembles. The present approach requires only an instantaneous control eigenframe invariant under the dominant disorder operators and a CD correction maintaining adiabaticity, ingredients routinely available across inhomogeneous atomic, trapped-ion, and solid-state platforms. Eigenframe engineering thus establishes a third axis for disorder-robust Floquet control, distinct from and complementary to quasienergy protection and average-Hamiltonian cancellation, which is applicable wherever collective coherence in a periodically driven ensemble is limited by static inhomogeneity.

\bibliography{apssamp}

\end{document}